\documentclass{ws-ijmpa}
\usepackage[super,compress]{cite}
\usepackage{url}
\usepackage[dvips,breaklinks]{hyperref}
\usepackage{breakurl}

\begin{document}

\title{PPN parameters in gravitational theory with nonminimally derivative coupling}

\author{Zhu Yi\footnote{Email: yizhuhust@sina.cn} \ and Yungui Gong\footnote{Email: yggong@mail.hust.edu.cn}}

\address{School of Physics, Huazhong
University of Science and Technology, Wuhan 430074, P.R. China}

\maketitle

\begin{abstract}
The nonminimal coupling of the kinetic term to Einstein's tensor helps the implementation of inflationary models
due to the gravitationally enhanced friction.
We calculate the parametrized post-Newtonian (PPN) parameters
for the scalar--tensor theory of gravity with nonminimally derivative coupling.
We find that under experimental constraint from
the orbits of millisecond pulsars in our galaxy,
the theory deviates from Einstein's general relativity in the order of $10^{-20}$, and
the effect of the nonminimal coupling is negligible if we take the scalar field as dynamical dark energy.
With the assumed conditions that the background scalar field is spatially homogeneous and evolves
only on cosmological timescales and the contribution to stress--energy in the solar system from
the background scalar field is subdominant, the scalar field is required to be massless.
\end{abstract}

\ccode{PACS numbers: 04.25.Nx, 04.50.Kd, 95.36.+x}

\section{Introduction}

The simplest generalization of Einstein's general relativity is the scalar--tensor theory of gravity such as Brans--Dicke theory \cite{Brans:1961sx}.
In Brans--Dicke theory, in addition to the massless spin two graviton,
a massless scalar field is also responsible for the exchange of gravitational interaction.
The scalar field called dilaton appears in the low energy effective bosonic string theory, and the scalar
degree of freedom arises naturally upon compactification of higher dimensions.
The most general scalar--tensor theory of gravity which gives at most the second--order equation of motion in four--dimensional spacetime is
Horndeski theory \cite{Horndeski:1974wa} with the following Lagrangian,
\begin{equation}
\label{horndeskieq1}
L_H=L_2+L_3+L_4+L_5,
\end{equation}
where
\begin{gather}
L_2=K(\phi,X),\quad L_3=-G_3(\phi,X)\Box \phi,\\
L_4=G_4(\phi,X)R+G_{4,X}\left[(\Box\phi)^2-(\nabla_\mu\nabla_\nu\phi)(\nabla^\mu\nabla^\nu\phi)\right],\\
\begin{split}
L_5=&G_5(\phi,X)G_{\mu\nu}\nabla^\mu\nabla^\nu\phi-\frac{1}{6}G_{5,X}\left[(\Box\phi)^3\right.\\
&\left.-3(\Box\phi)(\nabla_\mu\nabla_\nu\phi)(\nabla^\mu\nabla^\nu\phi)
+2(\nabla^\mu\nabla_\alpha\phi)(\nabla^\alpha\nabla_\beta\phi)(\nabla^\beta\nabla_\mu\phi)\right],
\end{split}
\end{gather}
$X=-\nabla_\mu\phi\nabla^\mu\phi/2$, $\Box\phi=\nabla_\alpha\nabla^\alpha\phi$,
the functions $K$, $G_3$, $G_4$ and $G_5$ are arbitrary functions of $\phi$ and $X$, and $G_{4,X}(\phi,X)=dG_4(\phi,X)/dX$.
The theory not only includes the
nonminimal coupling $f(\phi) R$, but also includes the nonminimal coupling between the kinetic term $X$ and $R$.
For the special nonminimal coupling $-\phi^2 R/6$ \cite{Callan:1970ze}, the theory is conformal invariant.
In general, there is no conformal invariance.

However, if we only consider the nonminimal interactions which are quadratic in $\phi$ and linear in $R$, then we only
need to consider the nonminimally derivative terms $(\nabla_\mu \phi \nabla^\mu \phi)R$, $(\nabla_\mu \phi \nabla_\nu \phi)R^{\mu\nu}$,
$\phi\Box\phi R$, $\phi(\nabla_\mu\nabla_\nu \phi) R^{\mu\nu}$, $\phi\nabla_\mu \phi \nabla^\mu R $ and $\phi^2\Box R$ \cite{Amendola:1993uh}.
Because of the divergencies of $\nabla_\mu (R\phi\nabla^\mu \phi)$, $\nabla_\mu (R^{\mu\nu}\phi\nabla_\nu \phi)$ and $\nabla_\mu (\phi^2 \nabla^\mu R)$,
the independent interactions are $(\nabla_\mu \phi \nabla^\mu \phi)R$, $(\nabla_\mu \phi \nabla_\nu \phi)R^{\mu\nu}$ and
$\phi\Box\phi R$. The inflationary solution was discussed in \cite{Amendola:1993uh} for the nonminimal coupling
$[\xi f(\phi)+\mu \nabla_\alpha\phi \nabla^\alpha\phi]R$, but the equation of motion for the scalar field is fourth order.
If the kinetic term is coupled to Einstein tensor by the coupling $\xi (\nabla_\mu\phi\nabla_\nu\phi) G^{\mu\nu}$,
then the field equations contain no more than second derivatives \cite{Capozziello:1999uwa,Sushkov:2009hk},
because it is the special
case of Horndeski theory of $L_5$ with $G_5(\phi,X)=\phi$ due to the divergence of $G^{\mu\nu}\nabla_\mu(\phi\nabla_\nu\phi)$.
Even for massless scalar field without a potential, the nonminimally derivative coupling $\xi (\nabla_\mu\phi\nabla_\nu\phi) G^{\mu\nu}$
has de-Sitter solution \cite{Capozziello:1999uwa}. If the canonical kinetic term $X$ is absent, the nonminimally derivative
coupling term behaves like a dark matter \cite{Gao:2010vr,Ghalee:2013smy}.
The introduction of the nonminimally derivative coupling to Einstein tensor causes the scalar
field to evolve more slowly because of the gravitationally enhanced friction, so the new Higgs
inflation was successfully implemented without either violating the unitarity bound or fine-tuning the
coupling constant $\lambda$ \cite{Germani:2010gm,Germani:2010ux,Germani:2011ua,Germani:2014hqa}.
Furthermore, the challenge to slow-roll inflation with single field due to the Lyth bound was also alleviated \cite{Yang:2015pga,Gong:2015p,Gao:2014pca,Gong:2014cqa}.
The cosmological consequences of the theory with nonminimally derivative coupling were discussed extensively \cite{Daniel:2007kk, Capozziello:1999xt,Tsujikawa:2012mk,Sadjadi:2013psa,Saridakis:2010mf,Sushkov:2012za,Skugoreva:2013ooa,
DeFelice:2011uc,Sadjadi:2010bz,Sadjadi:2013uza,Minamitsuji:2013ura,Granda:2009fh,
Granda:2010hb,Granda:2011eh,deRham:2011by,Jinno:2013fka,Sami:2012uh,Anabalon:2013oea,Rinaldi:2012vy,Koutsoumbas:2013boa,
Cisterna:2014nua,Huang:2014awa,Bravo-Gaete:2013dca,Bravo-Gaete:2014haa,Bruneton:2012zk,Feng:2013pba,
Feng:2014tka,Heisenberg:2014kea,Goodarzi:2014fna,Huang:2015yva,Huang:2016qkd,Cisterna:2015yla,Dalianis:2014sqa,Dalianis:2014nwa,Dalianis:2015aba,Ema:2015oaa,Aoki:2015eba,Harko:2015pma,Matsumoto:2015hua}.

We can also distinguish scalar--tensor theories of gravity from Einstein's general relativity by calculating the parametrized post-Newtonian (PPN) parameters
and compare them with the solar system experiments. For the original Brans--Dicke theory, the PPN parameter $\gamma=(1+\omega)/(2+\omega)$. If the scalar
field has a potential $V(\phi)$, then the PPN parameter $\gamma$ is \cite{Hohmann:2013rba}
\begin{equation}
\label{ppneq1}
\gamma(r)=\frac{2\omega_0+3-\exp(-m_\phi r)}{2\omega_0+3+\exp(-m_\phi r)},
\end{equation}
where the mass is
$$m_\phi=2\sqrt{\frac{8\pi G_N\phi_0}{2\omega_0+3}\frac{\partial^2V(\phi)}{\partial\phi^2}\Big|_{\phi_0}}.$$
The PPN analysis for the nonminimally derivative coupling $(\nabla_\mu\phi\nabla_\nu\phi)R^{\mu\nu}$
with massless scalar field was performed
in Ref. \refcite{Daniel:2007kk}. In this paper, we follow Ref. \refcite{Daniel:2007kk} to calculate the PPN parameters for
the scalar--tensor theory with the nonminimally derivative coupling $(\nabla_\mu\phi\nabla_\nu\phi)G^{\mu\nu}$.
In Sec. 2 we review the general equation of motion for the theory, and the PPN analysis was carried out in Sec. 3,
we conclude the paper in Sec. 4.


\section{Field Equations}
The action for the scalar--tensor theory of gravity with nonminimally derivative coupling to Einstein tensor is
\begin{equation}\label{action}
S=\int d^4x\sqrt{-g}\left\{\frac{R}{16\pi G}-\frac{1}{2}\left[g_{\mu\nu}-\omega^2 G_{\mu\nu}\right]\nabla^\mu \phi \nabla^\nu \phi-V(\phi)\right\}+S_m,
\end{equation}
where the nonminimally coupling constant $\omega^2$ has the dimension of length squared, $V(\phi)$ is the potential of the scalar field,
and $S_m$ is the action for the matter. Varying the
action \eqref{action} with respect to the metric $g_{\mu\nu}$, we get the gravitational field equation
\begin{equation}\label{gra:field:equation}
G_{\mu\nu}=8\pi G \left[T^{(m)}_{\mu\nu}+T^{(\phi)}_{\mu\nu}-\omega^2\Theta_{\mu\nu}\right],
\end{equation}
with $T^{(\phi)}_{\mu\nu}$
\begin{equation}\label{phi:tensor}
T^{(\phi)}_{\mu\nu}=\nabla_{\mu}\phi\nabla_{\nu}\phi-\frac12g_{\mu\nu}\left(\nabla\phi\right)^2-g_{\mu\nu}V(\phi),
\end{equation}
and $\Theta_{\mu\nu}$
\begin{equation}
\label{theta:tensor}
\begin{split}
\Theta_{\mu\nu}=&-\frac12\nabla_{\mu}\phi\nabla_{\nu}\phi R+2\nabla_\alpha\phi\nabla_{(\mu}\phi R_{\nu)}^{\alpha}\\
&-\frac12\left(\nabla\phi\right)^2G_{\mu\nu}+\nabla^{\alpha}\phi\nabla^{\beta}\phi R_{\mu\alpha\nu\beta}\\
&+\nabla_{\mu}\nabla^{\alpha}\phi\nabla_{\nu}\nabla_{\alpha}\phi-\nabla_{\mu}\nabla_{\nu}\phi\square\phi\\
&+g_{\mu\nu}\left[-\frac12\nabla^{\alpha}\nabla^{\beta}\phi\nabla_{\alpha}\nabla_{\beta}\phi+\frac12\left(\square\phi\right)^2
-\nabla_{\alpha}\phi\nabla_{\beta}\phi R^{\alpha\beta}\right].
\end{split}
\end{equation}
Then, we get
\begin{equation}\label{phi:stensor:s}
S^{(\phi)}_{\mu\nu}=\nabla_{\mu}\phi\nabla_{\nu}\phi+g_{\mu\nu}V(\phi),
\end{equation}
\begin{equation}\label{theta:tensor:s}
\begin{split}
S^{(\theta)}_{\mu\nu}=&-\frac12\nabla_{\mu}\phi\nabla_{\nu}\phi R+2\nabla_\alpha\phi\nabla_{(\mu}\phi R_{\nu)}^{\alpha}\\
&-\frac12\left(\nabla\phi\right)^2G_{\mu\nu}+\nabla^{\alpha}\phi\nabla^{\beta}\phi R_{\mu\alpha\nu\beta}\\
&+\nabla_{\mu}\nabla^{\alpha}\phi\nabla_{\nu}\nabla_{\alpha}\phi-\nabla_{\mu}\nabla_{\nu}\phi\square\phi\\
&-\frac12g_{\mu\nu}\nabla_{\alpha}\phi\nabla_{\beta}\phi R^{\alpha\beta},
\end{split}
\end{equation}
where $S_{\mu\nu}=T_{\mu\nu}-\frac12g_{\mu\nu}T$.
In terms of $S_{\mu\nu}$, we can express Einstein's field equation \eqref{gra:field:equation} as
\begin{equation}\label{gra:field:equation:s}
\frac{R_{\mu\nu}}{8\pi G}=S^{(m)}_{\mu\nu}+S^{(\phi)}_{\mu\nu}-\omega^2S^{(\theta)}_{\mu\nu}.
\end{equation}
Similarly, we derive the equation of motion for the scalar field as
\begin{equation}\label{sca:field:equation}
\left[g^{\mu\nu}-\omega^2 G^{\mu\nu}\right]\nabla_{\mu}\nabla_{\nu}\phi-V'(\phi)=0.
\end{equation}

\section{PPN Analysis}
In this section, we will calculate the metric in the solar system to the first--order beyond the Newtonian approximation.
The scalar field perturbation is written as $\delta\phi$.
The metric is expanded around the flat background $\eta_{\mu\nu}$ to the first--order, $g_{\mu\nu}=\eta_{\mu\nu}+h_{\mu\nu}$,
and the perturbation $h_{\mu\nu}$ is at least first--order in $(v/c)^2$, where $v$ is the characteristic velocity of bodies in the solar system.
The order of the Post-Newtonian expansion is determined by $GM/r\sim (v/c)^2$, where $O(1)\sim (v/c)^2$ and $O(1.5)\sim(v/c)^3$.
Our objective is to compute the metric $h_{\mu\nu}$ to the  order of $(v/c)^4$, or $O(2)$. Following Ref. \refcite{Daniel:2007kk},
we also make the following two assumptions: (C1) we assume that the background scalar field $\phi_0$ is spatially homogeneous and evolves
only on cosmological timescales which are much longer than the characteristic timescales of the solar system or laboratory, so
the spatial derivatives and second and higher time derivatives of $\phi_0$ are neglected, and
we only consider $\dot \phi_0$ in the second--order of $h_{00}$; (C2) we assume the contribution to stress--energy in the solar system from
the background scalar field is subdominant, so we can replace Ricci tensor in $S^{(\theta)}_{\mu\nu}$ by $8\pi G S^{(m)}_{\mu\nu}$.

To the zeroth--order, from Eq. \eqref{gra:field:equation:s} we get
\begin{equation}\label{zero:gra:field:equation}
\eta_{\mu\nu}V_0=0 \Rightarrow V_0=0,
\end{equation}
where $V_0=V(\phi_0)$.
To the first--order, Eq. (\ref{gra:field:equation:s}) becomes
\begin{equation}\label{pert:gra}
\begin{split}
\frac{\delta R_{\mu\nu}}{8\pi G}
&=\delta S^{(m)}_{\mu\nu}+\partial_{\mu}\delta\phi\partial_{\nu}\phi_0+\partial_{\mu}\phi_0\partial_{\nu}\delta\phi+\eta_{\mu\nu}V_1\delta\phi+h_{\mu\nu}V_0+\eta_{\mu\nu}\frac{V_2}{2}(\delta\phi)^2\\
& -\omega^2[-\frac12\partial_{\mu}\phi_0\partial_{\nu}\phi_0\delta R+2\dot{\phi_0}\nabla_{(\mu}\phi_0 \delta R^{0}_{\nu)}+\frac12\dot\phi_0^2\delta G_{\mu\nu}\\
&+\dot\phi_0^2\delta R_{\mu0\nu0}-\frac12\eta_{\mu\nu}\dot\phi_0^2\delta R^{00}],
\end{split}
\end{equation}
where $V_1=V'(\phi_0)$ and $V_2=V''(\phi_0)$.
To the zeroth--order, Eq. (\ref{sca:field:equation}) gives
\begin{equation}\label{zero:sca:field:equation}
V_1=0.
\end{equation}
The linear perturbed expansion of Eq. (\ref{sca:field:equation}) is
\begin{equation}\label{pert:sca}
\eta^{\mu\nu}\left(\partial_{\mu}\partial_{\nu}\delta\phi-\delta\Gamma^{0}_{\mu\nu}\dot{\phi_0}\right)-V_2\delta\phi=0,
\end{equation}
so we get
\begin{equation}\label{delta:sca:equation}
\left(\nabla^2-V_2\right)\delta\phi=\ddot{\delta\phi}+(\delta\Gamma^{0}_{ii}-\delta\Gamma^{0}_{00})\dot{\phi_0}.
\end{equation}
Since the order of $\delta\Gamma^0_{\mu\nu}$ is at least $O(1.5)$, so
\begin{equation}\label{delta:sca:equation:know}
  \delta\phi\sim O(1.5).
\end{equation}

To solve Eq. \eqref{pert:gra}, we need to calculate the perturbed Ricci tensor. In order to simplify the calculation,  we use the gauge conditions
\begin{equation}
\label{gauge}
\partial_{\mu}h_{i}^{\mu}=\frac12\partial_ih\quad\partial_{\mu}h^{\mu}_{0}=-\frac12\partial_{0}h_{00}+\frac12\partial_{0}h.
\end{equation}
With the above gauge choice, the perturbed Ricci tensor is
\begin{equation}\label{R00:1}
\delta R_{00}(h)=-\frac{1}{2}\nabla^2h_{00},
\end{equation}
\begin{equation}\label{Rij:1}
\delta R_{ij}(h)=-\frac{1}{2}\nabla^2h_{ij}+\frac{1}{2}\frac{\partial^2 h_{ij}}{\partial t^2},
\end{equation}
\begin{equation}\label{R0i:1}
\delta R_{0i}(h)=-\frac{1}{2}\nabla^2h_{0i}-\frac{1}{4}\frac{\partial^2 h_{kk}}{\partial x^i \partial t}+\frac{1}{2}\frac{\partial^2 h_{ik}}{\partial x^k \partial t},
\end{equation}
\begin{equation}\label{R00:2}
\delta R_{00}(h^{2})=-\frac12\nabla^2h^{(2)}_{00}-\frac{1}{2}h^{ij}\frac{\partial^2 h_{00}}{\partial x^i \partial x^j}-\frac{1}{4}(\nabla h_{00})^2-\frac{1}{4}\frac{\partial h_{00}}{\partial x^k}\frac{\partial h_{jj}}{\partial x^k}+\frac{1}{2}\frac{\partial h_{jk}}{\partial x^j}\frac{\partial h_{00}}{\partial x^k},
\end{equation}
where $h_{00}^{(2)}\sim O(2)$, $h_{00}\sim O(1)$ and $h_{0i}\sim O(1.5)$. Throughout this paper, we only label second and higher order terms like $h_{00}^{(2)}$.
We know the stress--energy tensor of a perfect fluid is
\begin{equation}\label{perf:fluid:tensor}
\begin{split}
T^{00}&=\rho(1+\Pi+v^2+2U)+O(2.5),\\
T^{0j}&=\rho(1+\Pi+v^2+2U+p/\rho)v^i+O(3),\\
T^{ij}&=\rho v^iv^j(1+\Pi+v^2+2U+p/\rho)+p\delta^{ij}(1-2\gamma U)+O(3.5),
\end{split}
\end{equation}
so we get $S_{\mu\nu}$ as
\begin{equation}\label{per:S}
\begin{split}
  S_{00}&=\frac12\rho\left(1+\Pi+2v^2+2U-2h_{00}+3p/\rho\right)+O(2.5),\\
  S_{ij}&=\frac{\rho}{2}\delta_{ij}+O(1.5),\\
  S_{0i}&=-\rho v^j+O(2).
\end{split}
\end{equation}
Now we are ready to compute the Newtonian limit of the $h_{00}$ and $h_{ij}$. To do this we should collect all $O(1)$ terms in Eq. (\ref{pert:gra}).
From equation (\ref{delta:sca:equation:know}) we know $\delta\phi\sim O(1.5)$,
so the $00$ component of the right--hand side of equation (\ref{pert:gra}) is $\rho/2$ and
the $ij$ components of the right--hand side of equation (\ref{pert:gra}) is $\rho/2\delta_{ij}$,
to the order of $O(1)$, we get
\begin{gather}
\label{h001}
\frac12\nabla^2h_{00}=-4\pi G\rho,\\
\label{hij1}
\frac12\nabla^2h_{ij}=-4\pi G\rho\delta_{ij},\\
\label{hueq1}
\nabla^2U=-4\pi G\rho.
\end{gather}
Therefore, the solution is
\begin{equation}
\label{hsol111}
h_{00}=2U,\quad h_{ij}=2U\delta_{ij}.
\end{equation}
To the order of $O(1.5)$, the $0i$ components of equation (\ref{pert:gra}) becomes
\begin{equation}
-\frac{1}{2}\nabla^2h_{0i}-\frac{1}{4}\frac{\partial^2 h_{kk}}{\partial x^i \partial t}+\frac{1}{2}\frac{\partial^2 h_{ik}}{\partial x^k \partial t}=-8\pi G\rho v_i,
\end{equation}
Substituting the solution $h_{ij}$ in Eq. \eqref{hsol111} to the above equation, we get the solution
\begin{equation}\label{h0i}
h_{0i}=-\frac72 V_i-\frac12 W_i,
\end{equation}
where
\begin{equation}
V_j\equiv\int\frac{G\rho(\mathbf{x'},t)v'_j}{\mathbf{|x-x'|}}d^3x',\quad W_j\equiv\int\frac{G\rho(\mathbf{x'},t)\mathbf{v'}\cdot\mathbf{(x-x')}(x-x')_j}{\mathbf{|x-x'|}^3}d^3x'.
\end{equation}
So far the solutions are the same as those in general relativity.
Then we need to get the second order solution of $h_{00}$.
From equations \eqref{pert:gra} and \eqref{delta:sca:equation:know}, we can get the  $O(2)$ terms of the sources
\begin{equation}\label{S00:2}
\frac{R^{(2)}_{00}}{8\pi G}=S^{(m,2)}_{00}+6\omega^2\pi G\rho\dot{\phi_0}^2,
\end{equation}
By introducing the Post-Newtonian potential $\Phi_2$ \cite{will},
\begin{equation}
\nabla^2\Phi_2=-4\pi G\rho U,\quad(\nabla U)^2=\nabla^2(\frac12U^2-\Phi_2),
\end{equation}
then equation (\ref{R00:2}) for the Ricci tensor becomes
\begin{equation}\label{R00:2:U}
R_{00}^{(2)}=-\frac12\nabla^2\left(h_{00}^{(2)}+2U^2-8\Phi_2
\right).
\end{equation}
From equation (\ref{per:S}) we know that
\begin{equation}\label{Sm00:2}
S^{(m,2)}_{00}=\rho\left(\frac{\Pi}{2}-U+v^2+\frac32p/\rho\right).
\end{equation}
Using the Post-Newtonian potentials \cite{will}
\begin{equation}\label{PPN:potential}
\nabla^2\Phi_1=-4\pi\rho Gv^2,\quad \nabla^2\Phi_3=-4\pi G\rho\Pi,\quad \nabla^2\Phi_4=-4\pi G p,
\end{equation}
we get
\begin{equation}\label{Sm00:2U}
S^{(m,2)}_{00}=-\frac{1}{8\pi G}\nabla^2\left(\Phi_3-2\Phi_2+2\Phi_1+3\Phi_4\right).
\end{equation}
Combining Eqs. \eqref{S00:2}, \eqref{R00:2:U} and \eqref{Sm00:2U}, we get
\begin{equation}
-\frac12\nabla^2\left(h_{00}^{(2)}+2U^2-8\Phi_2\right)=-\nabla^2\left(\Phi_3-2\Phi_2+2\Phi_1+3\Phi_4\right)-12\omega^2\pi G \nabla^2U\dot{\phi_0}^2,
\end{equation}
the solution is
\begin{equation}\label{metric:h}
h_{00}^{(2)}=24\pi G\omega^2\dot{\phi_0}^2 U-2U^2+2\Phi_3+4\Phi_2+4\Phi_1+6\Phi_4.
\end{equation}
Using the effective Newtonian constant and the dimensionless variable $X$,
$$G_{eff}=G\left(1+12\omega^2\pi G\dot{\phi_0}^2\right),\quad X=12\omega^2\pi G\dot{\phi_0}^2,$$
finally we get
\begin{equation}\label{metric:eff:h}
h_{00}=2U-\frac{2U^2}{(1-X)^2}+\frac{2\Phi_3}{1-X}+\frac{4\Phi_2}{(1-X)^2}+\frac{4\Phi_1}{1-X}+\frac{6\Phi_4}{1-X}+O(2.5),
\end{equation}
\begin{equation}
h_{ij}=\frac{2U\delta_{ij}}{1-X}+O(1.5),
\end{equation}
\begin{equation}
h_{0i}=-\frac{\frac72V_{i}}{1-X}-\frac{\frac12W_{i}}{1-X}+O(2),
\end{equation}
or
\begin{equation}
\begin{split}
g_{00}=-1+2U-2(1+2X)U^2+4(1+X)\Phi_1+4(1+2X)\Phi_2\\
+2(1+X)\Phi_3+6(1+X)\Phi_4,
\end{split}
\end{equation}
\begin{equation}
g_{ij}=\delta_{ij}+2U\delta_{ij}(1+X),
\end{equation}
\begin{equation}
g_{0i}=-\frac72V_{i}(1+X)-\frac12W_{i}(1+X).
\end{equation}
Comparing our results with the definitions of the PPN parameters \cite{will},
\begin{equation}\label{will:parameter}
\begin{split}
g_{00}=&-1+2U-2\beta U^2-2\xi\Phi_{W}+(2\gamma+2+\alpha_{3}+\zeta_{1}-2\xi)\Phi_{1}\\
&+2(3\gamma-2\beta+1+\zeta_2+\xi)\Phi_2+2(1+\zeta_3)\Phi_3\\
&+2(3\gamma+3\zeta_4-2\xi)\Phi_4-(\zeta_1-2\xi)A,
\end{split}
\end{equation}
\begin{equation}
g_{0i}=-\frac12(4\gamma+3+\alpha_1-\alpha_2+\zeta_1-2\xi)V_{i}-\frac12(1+\alpha_2-\zeta_1+2\xi)W_{i},
\end{equation}
\begin{equation}
g_{ij}=(1+2\gamma U)\delta_{ij}.
\end{equation}
then we have
\begin{equation}
\label{PPN:Parameter}
\begin{split}
&\gamma=1+X,\quad\beta=1+2X,\\
\xi=0,\quad&\alpha_1=4X,\quad\alpha_2=X\quad\alpha_3=2X,\\
\zeta_1=0,\quad&\zeta_2=5X,\quad\zeta_3=X,\quad\zeta_4=0.\\
\end{split}
\end{equation}
Note that in Einstein's general relativity
$\gamma=1$, $\beta=1$ and all other PPN parameters are zero.
Although the results are similar to that in Ref. \refcite{Daniel:2007kk}, the nonminimal coupling $\nabla_\mu\phi\nabla_\nu\phi R^{\mu\nu}$
considered there introduces extra degree of freedom, the coupling $\nabla_\mu\phi\nabla_\nu\phi G^{\mu\nu}$ considered
in this paper introduces no extra degree of freedom and the equation of motion is still the second--order differential equation.
By using the experimental constraints on the PPN parameters shown in Table \ref{tab:t1} \cite{Will:2014}, we get the current limit on $X$.
The current stringent limit on $X$ comes from
the constraint on the PPN parameter $|\alpha_3|\le 4\times 10^{-20}$ at the
95\% confidence level\cite{Bell:1996ir,Stairs:2005hu}, so $X\le 2\times 10^{-20}$ at the 95\% confidence level.

\begin{table}[htp]
\tbl{current limits on the PPN parameters \cite{Will:2014}.}
{\begin{tabular}{@{}ccc@{}} \toprule
Parameter&Observational Limit&Constraint on $X$\\\colrule
$\gamma-1$&$(2.1\pm 2.3)\times10^{-5}$ ~\cite{Bertotti:2003rm} & $(2.1\pm 2.3)\times10^{-5}$\\
$\beta-1$&$(1.34 \pm 0.043)\times10^{-5}$ ~\cite{Verma:2013ata} & $(6.7 \pm 0.2)\times10^{-6}$\\
$\alpha_1$&$4\times 10^{-5}$ ~\cite{Shao:2012eg} & $1\times 10^{-5}$\\
$\alpha_2$&$<1.6\times10^{-9}$ ~\cite{Shao:2013wga} & $<1.6\times10^{-9}$\\
$\alpha_3$&$\le 4\times10^{-20}$ ~\cite{Bell:1996ir,Stairs:2005hu} & $\le 2\times10^{-20}$\\
$\zeta_2$&$<4\times10^{-5}$& $<8\times 10^{-6}$\\
$\zeta_3$&$<1\times 10^{-8}$& $<10^{-8}$\\ \botrule
\end{tabular} \label{tab:t1}}
\end{table}

\section{Conclusions}

Horndeski theory is the most general scalar--tensor theory of gravity which introduces no extra degree of freedom because the equation
of motion of the theory contains at most the second derivatives. We consider the special case of Horndeski
theory in which the kinetic term of the scalar field is nonminimally coupled to Einstein tensor $G_{\mu\nu}$,
and we calculated the PPN parameters in this theory. The results we obtained are similar to those in Ref. \refcite{Daniel:2007kk}. However,
due to the nonminimal coupling of the kinetic term of the scalar field to Ricci tensor $R_{\mu\nu}$,
the theory considered in Ref. \refcite{Daniel:2007kk} introduces extra degree of freedom. Comparing with the
results in Einstein's general relativity, the PPN parameters have small corrections from $\dot\phi_0$ which is only important on the cosmological timescale.
Under the assumed conditions (C1) and (C2), the consistent solution requires that both $V_0$ and $V_2$ are zero, i.e., the scalar
field should be massless.

By using the constraint on the PPN parameter $\alpha_3\le 4\times 10^{-20}$ derived from the statistics of the period derivatives
of 21 millisecond pulsars \cite{Stairs:2005hu}, we find
the strongest limit $X=12\pi G\omega^2\dot\phi_0^2\le 2\times 10^{-20}$, so the theory with nonminimally derivative coupling
deviates from Einstein's general relativity in the order of $10^{-20}$.
The first time derivative of the orbital period of a selected sample of binary pulsars was also applied to test $f(R)$ theory of gravity \cite{DeLaurentis:2013zv}.

The ratio of the nonminimal kinetic term to the canonical one
is $3\omega^2 H_0^2$, so the nonminimal term becomes important only if $\omega>(\text{meV})^{-1}$ and the nonminimal term is negligible if
we choose $\omega \sim (10^{12}\text{GeV})^{-1}$ from inflationary models \cite{Yang:2015pga,Gong:2015p}.
Note that the contribution of the nonminimal coupling to the critical energy density of the Universe is $X$, so
the effect of the nonminimal coupling is negligible if the scalar field is taken as dynamical dark energy.
If we use the observational constraint $1+w\sim 0.2$ and $\Omega_m=0.3$ \cite{Gao:2013pfa,Gong:2013bn}, then we get the contribution of the
canonical kinetic energy is $(1+w)(1-\Omega_m)/2\sim 0.07$, so roughly we get $\omega^2 H_0^2 \sim 10^{-19}$ and the high friction limit assumed in
inflationary models \cite{Yang:2015pga,Gong:2015p} is guaranteed unless the equation of state parameter $w$ is extremely close to be -1.

\section*{Acknowledgements}
This research was supported in part by the Natural Science
Foundation of China under Grant No. 11475065,
and the Program for New Century Excellent Talents in University under Grant No. NCET-12-0205.


\end{document}